\documentclass[conference]{IEEEtran}
\IEEEoverridecommandlockouts
\usepackage{cite}
\usepackage{amsmath,amssymb,amsfonts}

\usepackage{amsthm}
\theoremstyle{definition}

\usepackage{algorithm,algorithmic}
\usepackage{booktabs}
\usepackage{graphicx}
\usepackage{textcomp}
\usepackage{caption}
\usepackage{subcaption}

\usepackage{xcolor}
\def\BibTeX{{\rm B\kern-.05em{\sc i\kern-.025em b}\kern-.08em
    T\kern-.1667em\lower.7ex\hbox{E}\kern-.125emX}}
\begin{document}

\title{Self-Sovereign Identity as a Service: \\ Architecture in Practice \\
\thanks{This is the preprint version of the conference paper "Self-Sovereign Identity as a Service: Architecture in Practice" in \textit{2022 IEEE 46th Annual Computers, Software, and Applications Conference (SAPSE 2022 Workshop)}.}
}

\author{\IEEEauthorblockN{Yepeng Ding}
\IEEEauthorblockA{
% \textit{dept. name of organization (of Aff.)} \\
\textit{The University of Tokyo}\\
Tokyo, Japan \\
youhoutei@satolab.itc.u-tokyo.ac.jp}
\and
\IEEEauthorblockN{Hiroyuki Sato}
\IEEEauthorblockA{
% \textit{dept. name of organization (of Aff.)} \\
\textit{The University of Tokyo}\\
Tokyo, Japan \\
schuko@satolab.itc.u-tokyo.ac.jp}
}
\maketitle

\begin{abstract}
Self-sovereign identity (SSI) has gained a large amount of interest. It enables physical entities to retain ownership and control of their digital identities, which naturally forms a conceptual decentralized architecture. With the support of the distributed ledger technology (DLT), it is possible to implement this conceptual decentralized architecture in practice and further bring technical advantages such as privacy protection, security enhancement, high availability. However, developing such a relatively new identity model has high costs and risks with uncertainty. To facilitate the use of the DLT-based SSI in practice, we formulate Self-Sovereign Identity as a Service (SSIaaS), a concept that enables a system, especially a system cluster, to readily adopt SSI as its identity model for identification, authentication, and authorization. We propose a practical architecture by elaborating the service concept, SSI, and DLT to implement SSIaaS platforms and SSI services. Besides, we present an architecture for constructing and customizing SSI services with a set of architectural patterns and provide corresponding evaluations. Furthermore, we demonstrate the feasibility of our proposed architecture in practice with Selfid, an SSIaaS platform based on our proposed architecture.
\end{abstract}


\section{Introduction}
\begin{figure}[htbp]
\centerline{\includegraphics[scale=0.39]{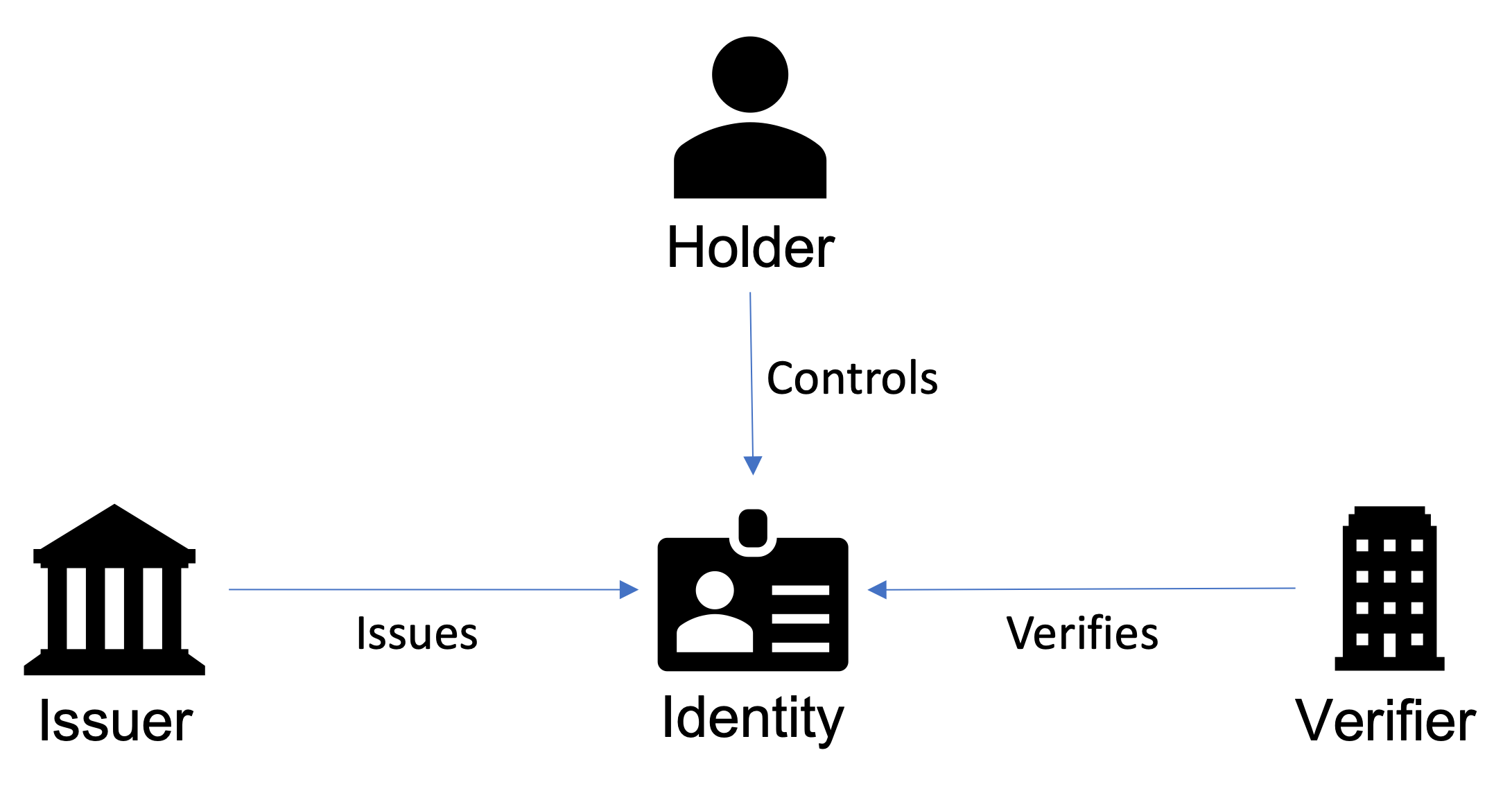}}
\caption{Conceptual diagram of self-sovereign identity (SSI).}
\label{fig:ssi}
\end{figure}

Self-sovereign identity (SSI) uses a user-centric mechanism that enables users to retain ownership of their identities and have full control of using them. Besides, users cannot append, modify, or remove any identity information without being detected. Hence, we conventionally call \textit{users} as \textit{holders}. Authorities, also called \textit{issuers}, play the role of issuing, updating, and revoking identities. Third parties requesting to verify the issued identities are called \textit{verifiers}. We depict a conceptual diagram in Figure~\ref{fig:ssi}. In SSI, holders obtain endorsed credentials, i.e., a set of claims associated with their identities, from issuers. Whenever holders are required to prove their identities or credentials by verifiers, they select and present a subset of credentials with proofs to verifiers after proving their identities. With or without directly interacting with issuers, verifiers can verify the validity of presented credentials based on proofs. All types of participants are centered around identities owned and controlled by holders. Under the assumption of the cryptographic verifiability of identities and associated information, the only trust in SSI is that verifiers trust issuers managing identities in an impartial manner. Therefore, it naturally forms a decentralized architecture. However, SSI has not been completely standardized and requires to be optimized for different scenarios in practice. Although the conceptual aspects of SSI have been intensively discussed, the architectural aspects and implementations are still in infancy.

Distributed ledger technology (DLT) has been widely used to enhance the security and privacy of systems in many fields such as finance \cite{chen_blockchain_2020}, Internet of Things \cite{dai_blockchain_2019}, data engineering \cite{ding_dagbase_2020}, and smart health \cite{ding_derepo_2020} on account of its decentralization characteristics. It enables immutable data persistence and processing without central entities. Particularly, there has been a growing interest in the adoption of blockchain technology, a type of DLTs, since the launch of the first cryptocurrency named Bitcoin. Furthermore, the blockchain has been put into industrial production as a service to facilitate the use of DLTs in practice, which is called Blockchain as a Service (BaaS). According to specific demands, both permissionless blockchains and permissioned blockchains are available through the service. Although the permissioned blockchain service is usually based on cloud infrastructure, it still improves the security of internal business services by decentralizing the conventional central entities, preventing the single point of failure, and ensuring integrity. Therefore, the adoption of DLTs, especially blockchain technology, has become a remarkable architectural decision to construct decentralized systems, including SSI systems.

However, the application of the DLT-based SSI has not been widely accepted in the industry due to its architecture risk and high implementation complexity. Motivated by standardizing the architecture of the DLT-based SSI, advocating its application in practice, and making it adapt to various scenarios with minimized efforts, we formulate and practicalize Self-Sovereign Identity as a Service (SSIaaS), a concept that promotes SSI to a service level, to facilitate the interoperable, customizable and scalable use of SSI as an identity model in systems. We summarize our main contributions as follows.

\begin{itemize}
    \item We propose a novel SSIaaS architecture that elaborates the service concept, SSI, and DLT for the implementation of SSIaaS platforms;
    \item We formulate a technical architecture for SSI services with a collection of architectural patterns for SSI service components and evaluate them in detail to provide insights on architectural decisions;
    \item We show Selfid, an SSIaaS platform based on our architecture to demonstrate the feasibility in practice.
\end{itemize}

\section{Related Work}
\label{sec:related_work}
DLTs have been used to support decentralized authentication and authorization of digital identities in recent years. In \cite{liu_identity_2017}, a blockchain-based IAMs is proposed to provide identity authentication and reputation management functionalities. In the context of the Internet of Things, an identity authentication scheme \cite{cui_hybrid_2020} is proposed for multi-WSN based on a hybrid blockchain model consisting of a private blockchain and a public blockchain. Bloccess \cite{ding_bloccess_2020} is proposed as an authorization framework based on a consortium blockchain to provide strong access control for distributed systems.

Besides, significant efforts have been made to implement SSI based on DLTs. For instance, Sovrin \cite{tobin_inevitable_2016} is a public and permissioned identity network operational on the Hyperledger Indy. It implements selective disclosure based on zero-knowledge proof (ZKP) techniques to enhance privacy. The SSI model in Sovrin is implemented based on permissioned blockchains. Therefore, although it is open to the public, its trust is addressed by the reputations and non-collusion of nodes.

Different from Sovrin, uPort \cite{naik_uport_2020} is an identity management system based on Ethereum and IPFS. uPort identities are implemented by smart contracts deployed on the Ethereum while the credentials and profile data are stored on the IPFS. And a \textit{Registry Contract} offers a cryptographic link between an on-chain identifier and its corresponding off-chain data. Furthermore, uPort provides wallets for users to facilitate the management of their identities.

From these SSI implementations, we find that a higher level can be abstracted by deconstructing them and aggregating functionally similar components only with different patterns. After the abstraction, we can regard each implementation as a service.

For the architectural perspective of services, BaaS is analyzed in a comprehensive way ranging from techniques to trust considerations in \cite{singh_blockchain_2018}. And in \cite{zheng_nutbaas_2019}, NutBaas is proposed and developed as a BaaS platform that provides blockchain services over cloud computing environments such as network deployment, system monitoring, as well as smart contracts analysis and testing.

To the best of our knowledge, self-sovereign identity as a service has not been systematically studied. Our work formulates this concept and focuses on its architectural aspect.

\section{Self-Sovereign Identity as a Service}
\label{sec:ssiaas}

\subsection{Overview}
\begin{figure*}[htbp]
\centerline{\includegraphics[scale=0.41]{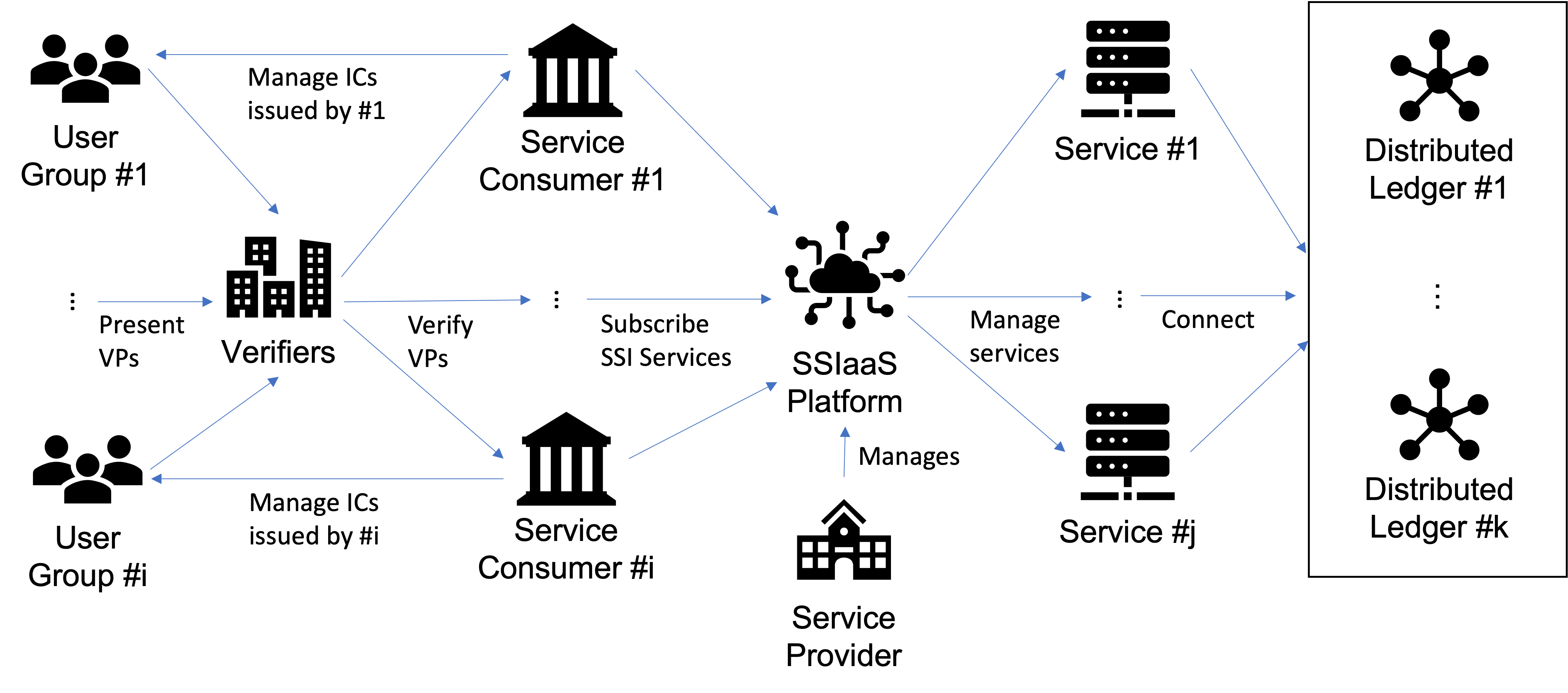}}
\caption{The conceptual architecture of self-sovereign identity as a service (SSIaaS).}
\label{fig:ssiaas_concept}
\end{figure*}

SSIaaS encapsulates SSI-based functionalities as a service and presents as a platform that provides application programming interfaces (APIs) for external invocation. The conceptual architecture of SSIaaS is depicted in Figure~\ref{fig:ssiaas_concept}.

There are four types of roles: service provider, service consumer, user, and verifier. Different from the conventional service providers, SSIaaS providers may not deploy all services in their cloud infrastructures. In extreme cases, they simply act as mediators who provide middlewares to facilitate interactions between service consumers and distributed ledgers hosting actual services. Service consumers are subscribers of SSI services. Although the concrete business model may affect the interaction between providers and consumers, it is out of the scope of this paper. Notably, service consumers in SSIaaS can also be service providers providing other types of services for their user groups. Users can be regarded as consumers of the services provided by service consumers in SSIaaS. The mapping from user group set to user set is surjective, i.e., each user belongs to at least one user group served by its corresponding service consumer. Nevertheless, the mappings from user group set to service consumer set and from service consumer set to service set are bijective. Hence, a service consumer can only manage the identities and credentials of its served user group. Verifiers play the same role as in SSI to verify credentials presented by users.

As we can see in Figure~\ref{fig:ssiaas_concept}, running services are managed by a service provider through an SSIaaS platform and interact with distributed ledgers. A service is more like a gateway and a front-end, while a distributed ledger is the actual execution environment for SSI functionalities. Therefore, the service provider and platform are not always necessary for service consumers to use SSI functionalities, though they seem to be in the central place of all relations. According to the concrete scenario, a service needs either a permissionless distributed ledger or a permissioned distributed ledger. Hence, the relation between service set and distributed ledger set is general, i.e., non-surjective and non-injective. For instance, a public blockchain, as a type of permissionless distributed ledger, can theoretically host a finite set of services.

\subsection{Self-Sovereign Identity Architecture}
\label{sec:ssi_architecture}
\begin{figure}[htbp]
\centerline{\includegraphics[scale=0.39]{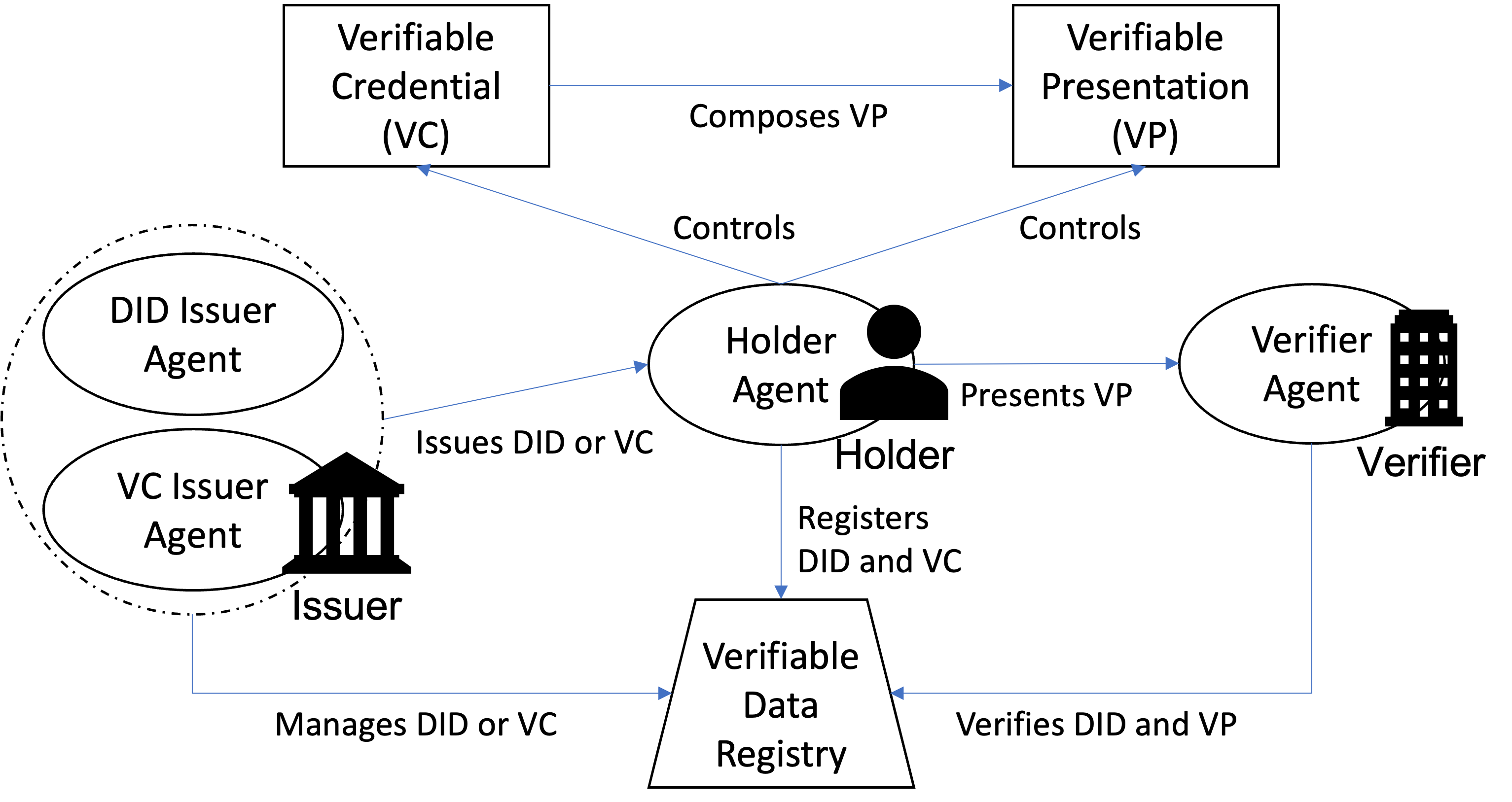}}
\caption{Self-sovereign identity architecture for service construction.}
\label{fig:ssi_architecture}
\end{figure}

We formulate a technical SSI architecture for service construction that is depicted in Figure~\ref{fig:ssi_architecture}. The architecture contains all three roles of SSI: holder, issuer, and verifier. Technically, a holder is represented by an agent that facilitates registering decentralized identifiers (DIDs) and verifiable credentials (VCs), requesting VCs from issuers, storing VCs, composing verifiable presentations (VPs), and presenting VPs to verifiers. Therefore, a holder agent acts as a client. Similarly, an issuer (or verifier) agent, either a DID issuer agent or a VC issuer agent, provides interfaces for issuers (or verifiers) to interact with holder agents and the verifiable data registry (VDR), which also acts as a client. Besides, a VDR is responsible for creating, modifying, revoking, and verifying DIDs and VCs, which acts as a server. It is notable that if the VDR is controlled by the issuer with a centralized mechanism, then this architecture can be regarded as a variant of the client-server architecture, which will not be discussed further in this paper. Mainly, public-key cryptography is used to generate and verify proofs of VCs.

We can map concepts in this architecture to concepts in SSIaaS. A user is abstracted as a holder who controls VCs issued by service consumers via a holder agent. A service consumer not only has issuer functionalities to issue VCs but also has the additional responsibility to link DIDs to corresponding physical identities via an issuer agent. The actual execution of the management functions is handled by the subscribed running service that provides full functionalities of the VDR. The concepts of the verifier in SSI and SSIaaS are equivalent. A verifier agent authenticates users, requests VPs, and verifies them through the VDR on behalf of a verifier.

Therefore, we can build various services based on this SSI architecture by customizing the technical parts: holder agent, issuer agent, verifier agent, and VDR.

\subsection{Service Architecture}
\label{sec:service_architecture}
\begin{figure}[htbp]
\centerline{\includegraphics[scale=0.39]{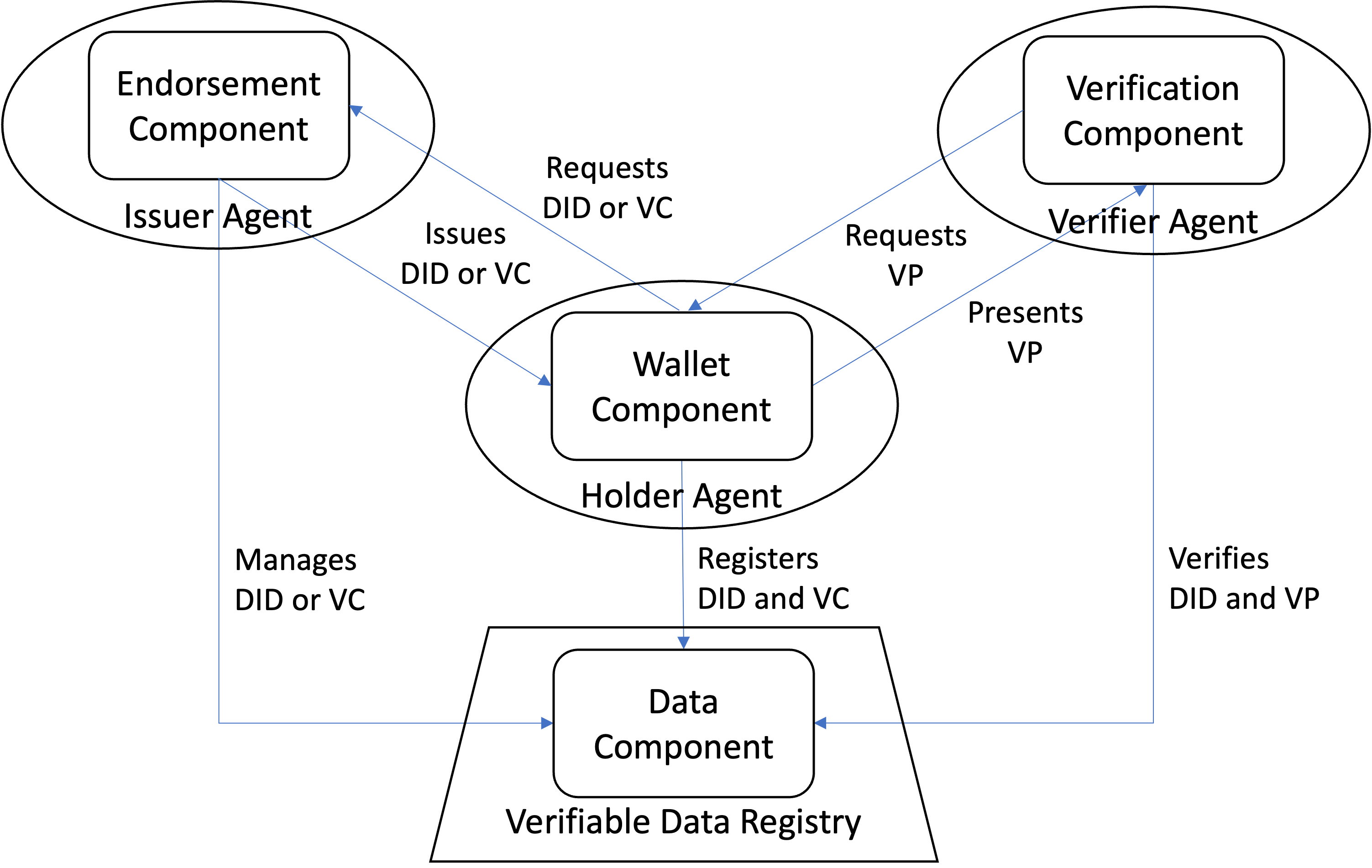}}
\caption{Self-sovereign identity service architecture.}
\label{fig:ssi_service_architecture}
\end{figure}

An SSI service is a minimal element in SSIaaS, which serves a service consumer and is uniformly managed by the service provider. As shown in Figure~\ref{fig:ssi_service_architecture}, an SSI service consists of four components: Data Component ($\mathfrak{D}$), Wallet Component ($\mathfrak{W}$), Endorsement Component ($\mathfrak{E}$), and Verification Component ($\mathfrak{V}$). Each component corresponds to a technical part of the SSI architecture illustrated in Section~\ref{sec:ssi_architecture}.

\subsubsection{DID Registration Protocol}
This protocol implements that a user $U$ with an identity $I_u$ obtains a DID $D_u$ from a service consumer $S$ and registers $D_u$ in $\mathfrak{D}$.

\begin{enumerate}
    \item $U$ generates a pair of keys $(\textit{PK}_u, \textit{SK}_u)$ via $\mathfrak{W}$;
    \item $U$ sends proof documents that support $I_u$ and $\textit{PK}_u$ to $S$;
    \item If $I_u$ is approved, $S$ uses $\mathfrak{E}$ to generate $D_u$ together with the DID document $\mathcal{R}(D_u)$ that includes $\textit{PK}_u$, DID controller of $S$, and verification methods. Then $D_u$ is returned to $U$;
    \item $U$ uses $\mathfrak{W}$ to invoke $\textit{DIDRegistration}(D_u, [\textit{Sig}_u])$ function of $\mathfrak{D}$ through $\mathfrak{W}$, where $\textit{Sig}_u=\mathcal{S}(D_u)$ is a signature signed with $\textit{SK}_u$ and an optional parameter only needed for delegated registration purposes;
    \item $\mathfrak{D}$ authenticates $U$ by checking whether $U$ can recover challenging message $M$ from its encoded form $\mathcal{E}(M, \textit{PK}_u)$ or authenticates the DID subject in delegation scenarios by verifying $\textit{Sig}_u$ with $\textit{PK}_u$ in $\mathcal{R}(D_u)$;
    \item If the authentication succeeds, $\mathfrak{D}$ stores tuple $(D_u, \mathcal{H}(\mathcal{R}(D_u)))$ to the underlying distributed ledger. Then a finalized transaction is returned to $U$ and saved in $\mathfrak{W}$.
\end{enumerate}

\subsubsection{VC Registration Protocol}
This protocol is executed by a user $U$ with a credential $C_u$ to obtain a VC $\textit{VC}_u$ from a service provider $S$ and register $\textit{VC}_u$ in $\mathfrak{D}$.

\begin{enumerate}
    \item $U$ sends proof documents of $C_u$ to $S$;
    \item $\mathfrak{E}$ of $S$ authenticates $U$ by $\textit{PK}_u$ in $\mathcal{R}(D_u)$;
    \item If the authentication succeeds and $C_u$ is approved, $\mathfrak{E}$ generates $\textit{VC}_u$ with proof $P_u= \mathcal{P}(\mathcal{H}(\textit{VC}_u))$ signed with $\textit{SK}_s$. Then $\textit{VC}_u$ and $P_u$ are returned to $U$;
    \item $U$ uses $\mathfrak{W}$ to sign $\mathcal{H}(\textit{VC}_u)$ with $\textit{SK}_u$ and get a signature $\textit{Sig}_u$, and invoke $\textit{VCRegistration}(D_u, \textit{VC}_u, P_u, \textit{Sig}_u)$ function of $\mathfrak{D}$ through $\mathfrak{W}$;
    \item $\mathfrak{D}$ authenticates $U$ in the same manner as in the \textit{DID Registration Protocol}. Additionally, $\mathfrak{D}$ checks the equivalence of the authenticated $D_u$ and the DID in the VC subject to verify that $U$ is the subject of $\textit{VC}_u$;
    \item If the authentication succeeds, $\mathfrak{D}$ verifies the integrity of $\textit{VC}_u$ by checking the existence of $S$ and $P_u$ with $\textit{PK}_s$ obtained from $\textit{VC}_u$;
    \item If the integrity checking succeeds, $\mathfrak{D}$ stores $\mathcal{H}(\textit{VC}_u)$ to the underlying distributed ledger. Then a finalized transaction is returned to $U$ and saved in $\mathfrak{W}$.
\end{enumerate}

Any direct modification of $C_u$ in $\textit{VC}_u$ will be easily detected by checking $P_u$ under the trust of the cryptography. We discuss two cases that bypass the integrity verification: proof forgery and issuer masquerade. In the first case, $U$ tampers with $C_u$, $\textit{PK}_s$, and $P_u$ at the same time with a self-generated key pair $(\textit{PK}'_s, \textit{SK}'_s)$. However, $\textit{VC}_u$ will be rejected because $\textit{PK}'_s,$ points to an unrecognized service consumer. In the second case, $U$ uses a service consumer $S'$ to masquerade as $S$ to bypass the existence checking. Although the execution of the \textit{VC Registration Protocol} succeeds, verifiers can still detect the masquerade by the \textit{VC Verification Protocol} illustrated next and reject $\textit{VC}_u$ issued by a falsified $S$.

\subsubsection{VC Verification Protocol}
A verifier $V$ uses this protocol to verify a presented VC $\textit{VC}_u$ issued by a service provider $S$ and from a user $U$.

\begin{enumerate}
    \item $\mathfrak{V}$ of $V$ checks the status of $\textit{VC}_u$ by querying the existence of $\mathcal{H}(\textit{VC}_u)$ in the distributed ledger and confirming the non-existence of a revoking mark;
    \item If the status checking succeeds, $\mathfrak{V}$ authenticates $U$ by checking whether $U$ can recover challenging message $M$ from its encoded form $\mathcal{E}(M, \textit{PK}_u)$ where $\textit{PK}_u$ is obtained from the presented $\textit{VC}_u$;
    \item If the authentication succeeds, $\mathfrak{V}$ checks the equivalence of $\textit{PK}_s$ in $\textit{VC}_u$ and the one obtained from $S$ to verify that $\textit{VC}_u$ is issued by the actual $S$;
    \item If the existence checking succeeds, $\mathfrak{V}$ verifies the integrity of $\textit{VC}_u$ by checking $P_u$ with $\textit{PK}_s$;
    \item If the integrity checking succeeds, $V$ can admit the validity of $\textit{VC}_u$.
\end{enumerate}

For the existence checking of the \textit{VC Verification Protocol}, $\mathfrak{V}$ obtains the real $\textit{PK}_s$ by a name service, a module of service management controlled by the service provider, which will be illustrated in Section~\ref{sec:name_service}. Besides, $S$ can provide a double-check method by making $\textit{PK}_s$ public in a trustworthy manner (e.g., hosting an official web application).

\subsection{Service Architectural Patterns}
\label{sec:pattern}
All four components are decoupled but follow common interfaces and generalized protocols, which infers that any architectural changes within a component do not affect the others. Therefore, services can be constructed from all kinds of combinations of feasible architectures of these interoperable components, enabling services to have heterogeneous architectures while preserving the same SSI functionalities. We formulate and analyze a set of architectural patterns for each component and make all components customizable to adapt to a wide range of scenarios.

\subsubsection{Data Component}
As a fundamental component, $\mathfrak{D}$ is responsible for implementing functionalities of the VDR that facilitates to register, update, revoke, and query DIDs and VCs.

\begin{itemize}
    \item The registering functionalities are implementations of the \textit{DID Registration Protocol} and \textit{VC Registration Protocol} in Section~\ref{sec:service_architecture}. Data required to be stored in the distributed ledger such as hash values are included in emitted \textit{DID (VC) Registered Events}.
    \item The updating functionalities use similar protocols to store modified information about DIDs and VCs in the distributed ledger and emit \textit{DID (VC) Updated Events}. Since users cannot modify DIDs and VCs without being detected, the updating protocols are executed by service consumers to make authorized updates.
    \item The revoking functionalities are also used by service consumers to store revoking marks that indicate the invalidation of the target DIDs or VCs by emitting \textit{Revoked Events} in the distributed ledger.
    \item The query functionality is implemented based on a linear search protocol in timestamp-reverse order to return the latest information and status of the target DIDs and VCs.
\end{itemize}

We formulate three architectural patterns: permissionless pattern, permissioned pattern, and sub-ledger pattern. An SSIaaS platform supporting all types of patterns can distinguish them by the identifiers of DID methods.

\paragraph{Permissionless Pattern}
In this pattern, $\mathfrak{D}$ implements the VDR functionalities by interacting with a permissionless distributed ledger such as a public blockchain with a proof-based consensus mechanism. The abbreviation of the adopted blockchain defines the identifier for the DID method.

For instance, smart contracts that implement data and event persistence can be deployed on Ethereum during the initialization of $\mathfrak{D}$. In this manner, $\mathfrak{D}$ can use application binary interfaces (ABIs) to interact with these smart contracts to implement full functionalities of the VDR.

\paragraph{Permissioned Pattern}
This pattern constructs the VDR based on a permissioned distributed ledger such as a consortium blockchain or a private blockchain with a voting or multi-party consensus mechanism. Platform-specific smart contracts are implemented to support the same VDR functionalities. However, the CDL-based and PDL-based architectures have significant differences compared to the permissionless distributed ledger.

\begin{figure}[htbp]
     \centering
     \begin{subfigure}[b]{0.35\textwidth}
         \centering
         \includegraphics[width=\textwidth]{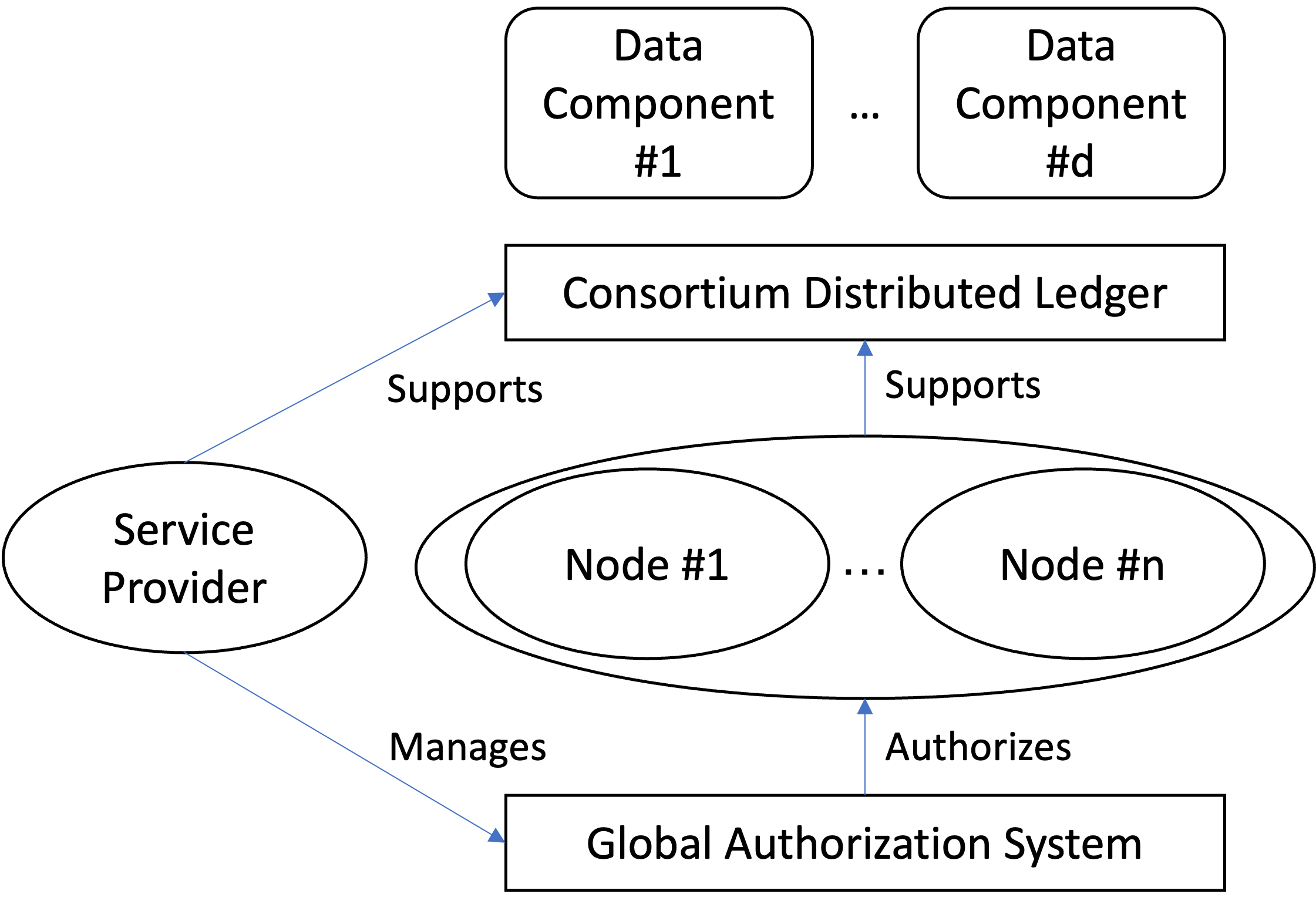}
         \caption{}
         \label{fig:pattern_consortium}
     \end{subfigure}
     \hfill
     \begin{subfigure}[b]{0.35\textwidth}
         \centering
         \includegraphics[width=\textwidth]{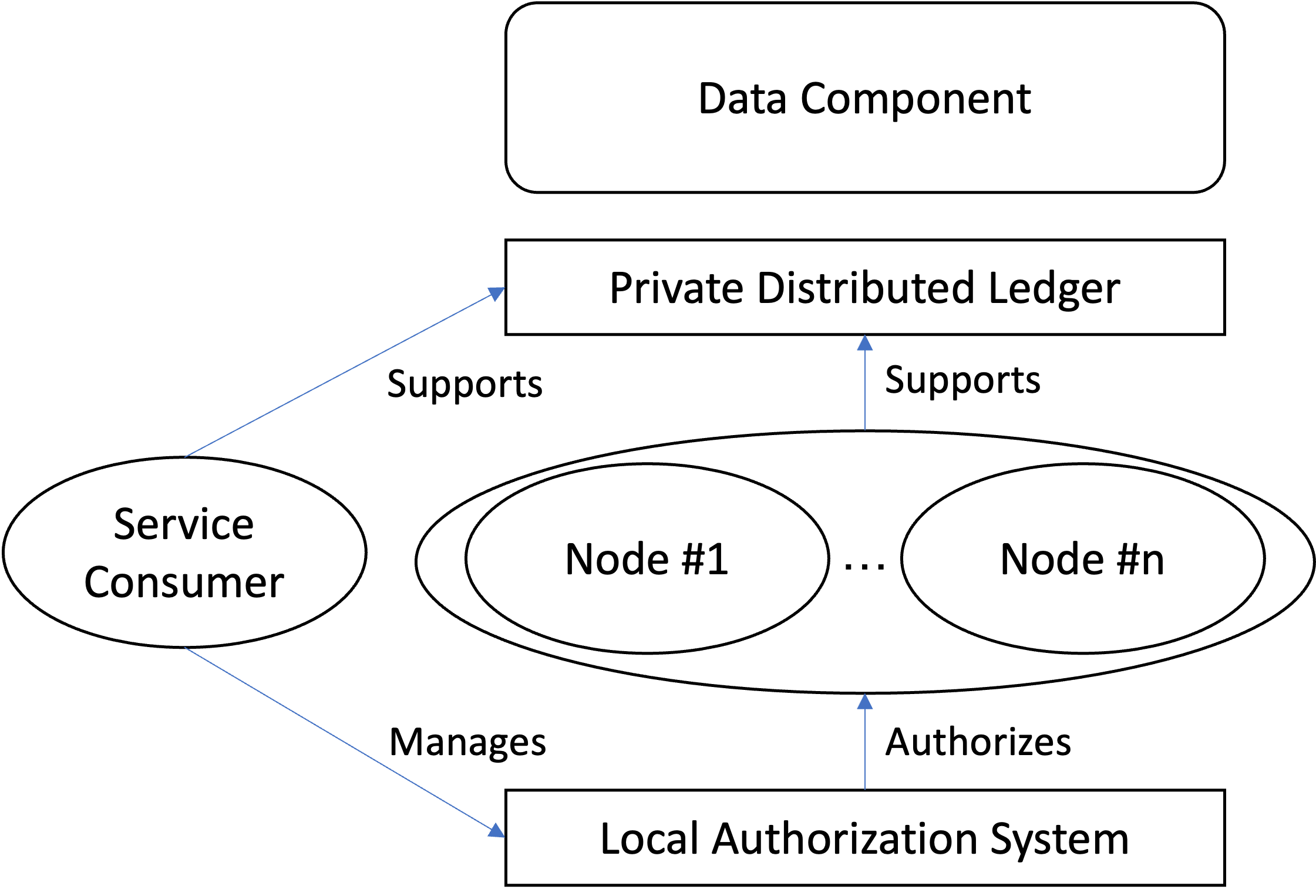}
         \caption{}
         \label{fig:pattern_private}
     \end{subfigure}
\caption{\textit{Permissioned Pattern} with a consortium ledger in Figure~\ref{fig:pattern_consortium} or a private distributed ledger in Figure~\ref{fig:pattern_private}.}
\end{figure}

For the adoption of a CDL, the service provider manages a global authorization system to authorize third parties to hold nodes that contribute computing power to the CDL, which is shown in Figure~\ref{fig:pattern_consortium}. This architecture implements the VDR for all provided SSI services based on one CDL. Since this pattern is mainly managed by the service provider, the identifier for the DID method is defined as the abbreviation of the name of the service provider.

If a service consumer chooses to use a PDL for a service, then the service provider allocates computing resources including a local authorization system and nodes. In other words, each SSI service has its unique VDR and authorization system, which is shown in Figure~\ref{fig:pattern_private}. The identifier for the DID method is defined as the abbreviation of the name of the service.

\paragraph{Sub-Ledger Pattern}
A sub-ledger is a slave distributed ledger that interacts with a permissionless distributed ledger as its master ledger. Sub-ledgers are usually permissioned distributed ledgers that afford a certain amount of computing work for their master chains. Consequently, this pattern has a hybrid architecture that inherits merits from both permissioned and permissionless distributed ledgers.

\begin{figure}[htbp]
\centerline{\includegraphics[scale=0.35]{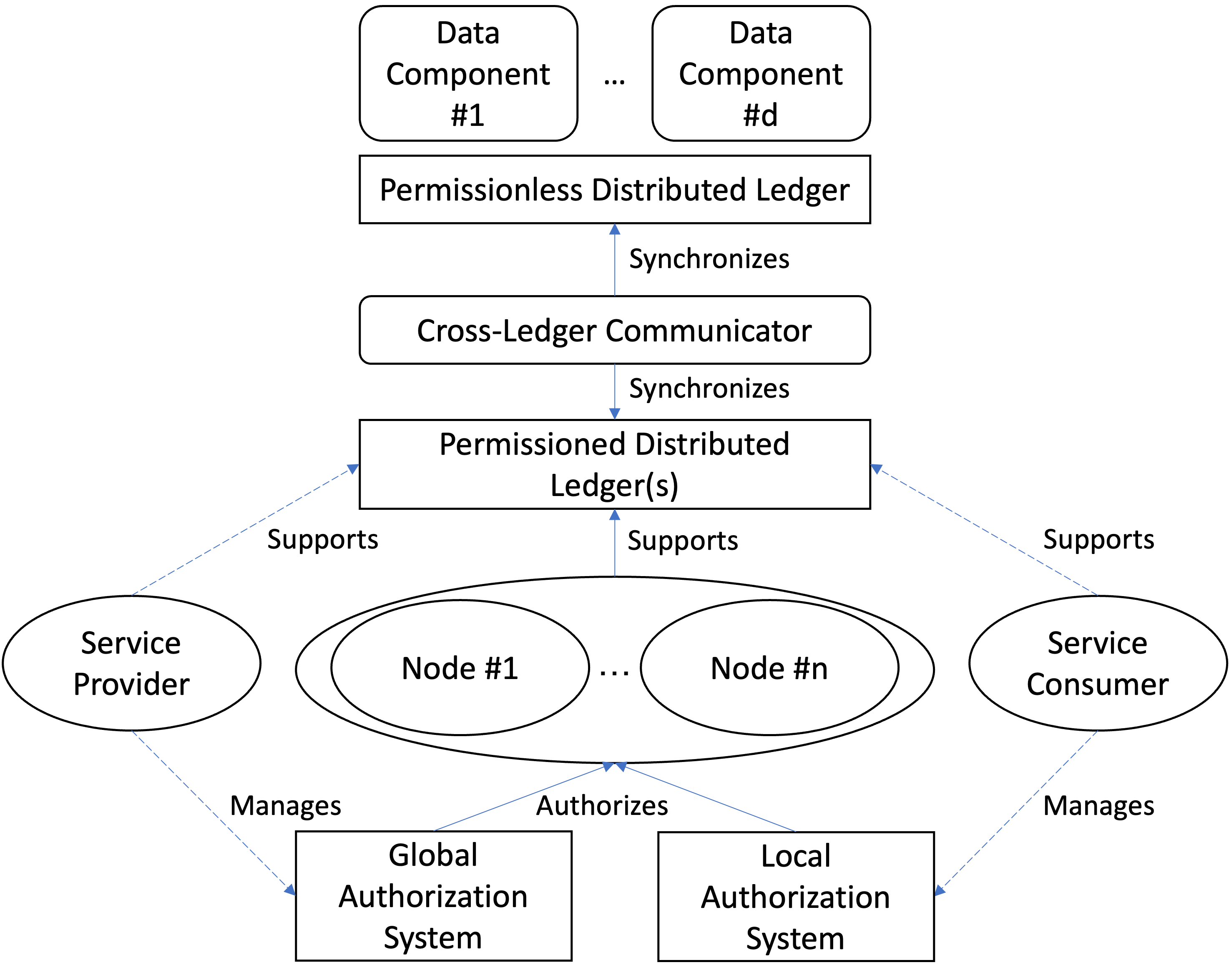}}
\caption{\textit{Sub-ledger pattern}.}
\label{fig:pattern_subledger}
\end{figure}

We depict this pattern in Figure~\ref{fig:pattern_subledger}. This pattern uses a sub-ledger as an interface for external interactions. The sub-ledger implements functionalities of the VDR in a similar way as the permissioned pattern. Besides, a cross-ledger communicator is responsible for packaging a set of operations enforced on the sub-ledger into a bundle and periodically storing its metadata on the master ledger.

\paragraph*{Evaluation of Patterns for Data Component}
The \textit{Permissionless Pattern} has better security properties, including integrity and availability, compared to the \textit{Permissioned Pattern} due to its proof-based consensus mechanisms that are fully decentralized and not controlled by service providers. However, these consensus mechanisms have low transaction throughput and lead to low efficiency of writing protocols of $\mathfrak{D}$ such as \textit{Registering}, \textit{Updating}, and \textit{Revoking}. Besides, permissionless distributed ledgers are usually large and contain vast amounts of irrelevant data, which causes low efficiency of reading protocols such as \textit{Querying}.

The \textit{Permissioned Pattern} makes trade-offs between security and efficiency. It has higher transaction throughput than the \textit{Permissionless Pattern}. Since the consortium blockchain is governed by service providers, it requires trust between the consumer and provider. As for the private blockchain, computing resources are provided by service consumers. Hence, service consumers have full control of the security of $\mathfrak{D}$. In each case, integrity and availability are potentially vulnerable due to the existence of centralized entities.

The \textit{Sub-Ledger Pattern} makes the balance between the \textit{Permissionless Pattern} and the \textit{Permissioned Pattern} by adopting a hybrid architecture. It improves the efficiency of the \textit{Permissionless Pattern} and enhances the integrity of the \textit{Permissioned Pattern}. However, it has lower availability than the \textit{Permissionless Pattern} and costs higher computing power consumption than the \textit{Permissioned Pattern}.

\subsubsection{Wallet Component}
$\mathfrak{W}$ is encapsulated into a holder agent. Its main function is to facilitate holders to manage DIDs and VCs including requesting, registering, composing VPs, and presenting VPs. We formulate two architectural patterns: online pattern and offline pattern. Both patterns share the common interface to interact with the other components, though their implementations are significantly different.

\paragraph{Online Pattern}
This pattern uses an online storage system to store a DID and VCs for a holder. With the holder's permission, $\mathfrak{W}$ stores the DID document and the obtained VCs to an online storage system, retrieves them to compose VPs, and presents them to the verifier on demand. An online storage system can be a cloud storage system or a decentralized storage system such as IPFS. 

Notably, both will be vulnerable without cryptographic measurements to protect security and privacy. A cloud storage system has a centralized mechanism that faces issues such as single point of failure and data tampering. A decentralized storage system is transparent, i.e., anyone with a correct link can retrieve the corresponding data from the system, which can cause privacy violations and credential leakage. Therefore, it requires cryptographic algorithms and carefully formulated protocols \cite{ding_sunspot_2021}. Two types of cryptographic algorithms are commonly available. Symmetric cryptography (e.g., Advanced Encryption Standard) can be used to generate a secret key for encryption. The key is locally protected by $\mathfrak{W}$, and only encrypted data are stored on an online storage system. By using asymmetric cryptography (e.g., Elliptic Curve Digital Signature Algorithm), a holder can use the public key to encrypt data and store them online while keeping the private key in $\mathfrak{W}$.

\paragraph{Offline Pattern}
The offline pattern uses a native management system that stores a DID and VCs for a holder. A holder has complete control of the system. Besides, the system can protect DIDs and VCs by introducing authentication mechanisms such as password authentication, public-key cryptography, and biometric authentication.

\paragraph*{Evaluation of Patterns for Wallet Component}
The \textit{Online Pattern} has good scalability, recoverability, and portability for both cloud-based storage and decentralized storage solutions. A cloud-based storage solution is more extensible than a decentralized storage solution because it can support other types of computing services besides storage. Nevertheless, a centralized mechanism lowers its availability and integrity. On the contrary, a decentralized storage solution has better security properties but lacks extensibility.

The \textit{Offline Pattern} shifts the security problem of protecting DIDs and VCs to holders. It has better privacy than the \textit{Online Pattern} but requires extra efforts to scale up and recover stored data. Without the backup functionality, it is impossible to recover data if the offline device is lost. Furthermore, the implementation of the native management system could be platform-specific, which leads to poor potability.

Notably, private key protection plays a critical role in authentication security in either case. If a private key associated with a DID is leaked, then the identity proved by that key can be stolen together with all credentials. Therefore, enabling multi-factor authentication in $\mathfrak{W}$ is a good option for service consumers to enhance authentication security. Service providers can use fast IDentity Online (FIDO) techniques to provide customizable multi-factor authentication functionalities.

\subsubsection{Endorsement Component}
$\mathfrak{E}$ serves DID issuers and VC issuers to facilitate issuing DIDs and VCs, as well as managing them. It functions partially when integrated into a \textit{DID Issuer Agent} or a \textit{VC Issuer Agent}. In the case where service consumers issue and manage both DIDs and VCs, $\mathfrak{E}$ works with full functionalities.

An issuer can consist of one or multiple entities. For the multi-entity case, the trust of an endorsement can be enhanced by a multi-review mechanism. A DID or a VC can only be issued when a set of entities have reviewed and approved it. We formulate two architectural patterns for the multi-entity case: multisignature pattern and secret-sharing pattern. For brevity, we take $\mathfrak{E}$ for VC issuers for example, while $\mathfrak{E}$ for DID issuers works in the same manner.

\paragraph{Multisignature Pattern}
This pattern implements a controller that dispatches a credential decoded from a VC request with proof documents to a set of randomly selected entities and collects their signatures. If all signatures are aggregated, the controller will sign a VC for the credential and notify $\mathfrak{W}$. Otherwise, the controller will reject the VC issuing process. This mechanism is based on a multisignature scheme \cite{gennaro_threshold-optimal_2016}. The signing power of VCs is shared amongst all entities. In a $t$-threshold scheme, a subset of $t + 1$ entities can jointly sign VCs, but a smaller subset cannot.

\paragraph{Secret-Sharing Pattern}
A secret-sharing scheme is implemented in this pattern, such as Shamir's Secret Sharing \cite{shamir_how_1979}. Assume an issuer consists of $n$ entities. In the initial stage, the controller encrypts the private key for signing VC with a secret key and divides the secret key into $n$ parts, and dispatches a random part to each entity. In a $(t, n)$-threshold scheme, to sign a VC for a credential with the encrypted private key, at least $t$ issuers have to review and approve that credential by submitting their parts of the key to the controller. By collecting $t$ parts of the secret key, the controller can reconstruct the private key to sign a VC.

\paragraph*{Evaluation of Patterns for Endorsement Component}
The \textit{Multisignature Pattern} protects the private key of signing VCs in a decentralized manner. All issuing entities hold and protect their individual private keys for signing. A VC is signed if and only if a certain amount of signatures are aggregated. Therefore, this pattern is resilient to key leakage. Nevertheless, the multisignature scheme is limited to specific digital signature algorithms (DSAs).

The \textit{Secret-Sharing Pattern} protects the private key of signing VCs by encryption. Each issuing entity holds a part of the secret key for decryption. Only when a certain amount of parts is collected, a VC can be signed with the decrypted private key reconstructed from collected parts. It can support various DSAs, but its security highly depends on the secret safety.

\subsubsection{Verification Component}
$\mathfrak{V}$ encapsulates interactions with $\mathfrak{W}$ and $\mathfrak{D}$ to facilitate verifiers to verify the validity of DIDs and VPs, which is an essential part of a verifier agent. We present two architectural patterns for $\mathfrak{V}$: service pattern and host pattern.

\paragraph{Service Pattern}
In this pattern, a service provider provides a verification service that implements functionalities of $\mathfrak{V}$. Service consumers can subscribe to the verification service and delegate functionalities of $\mathfrak{V}$ to the service provider. In this manner, verifiers verify obtained VPs via verification services provided by service providers.

\paragraph{Host Pattern}
Software development kits (SDKs) implemented with full functionalities of $\mathfrak{V}$ are distributed to service consumers in this pattern. Service consumers can develop a customized verification system based on the provided SDKs. The developed verification systems are used to serve verifiers to verify VPs.

\paragraph*{Evaluation of Patterns for Endorsement Component}
The \textit{Service Pattern} is ready-to-use for service consumers to provide verification functionalities for verifiers without development and computing resources. Correspondingly, service consumers have limited control of verification requests and can hardly extend functionalities.

On the contrary, the \textit{Host Pattern} gives service consumers full control of verification requests and flexibility of adding new functionalities such as authenticating verifiers. However, it requires service consumers to make extra efforts to develop and deploy verification systems based on SDKs.

\subsection{Service Management}
\label{sec:management}
From the perspective of service providers, an SSIaaS platform needs to support customizing service components, mapping requests to corresponding functionalities or services, monitoring service status. We illustrate four main modules including Parser, Gateway, Name Service, and Monitor.

\subsubsection{Parser}
As illustrated in Section~\ref{sec:pattern}, there are many kinds of service architectures derived from combinations of architectural patterns. To instantiate a customized service, a service consumer creates a specification file that contains a collection of selected architectural patterns for each component with concrete parameters in YAML format, and sends an encapsulated request to \textit{Gateway}.

After receiving a request from \textit{Gateway}, \textit{Parser} executes a parsing algorithm to generate a build program according to the parsed specification. Then the generated build program is executed to build all components, generate client and server applications based on the components, create execution environments (e.g., isolated networks and containers), deploy the server applications to the environments, and return the client applications.

\subsubsection{Gateway}
There are two types of requests: internal requests and external requests. Internal requests are management-level requests sent from service consumers and service providers, while external requests are service-level requests sent from users and verifiers. The \textit{Gateway} recognizes legitimate requests and dispatch them to corresponding functionalities or services.

The internal requests are sent to internal modules such as \textit{Parser} and \textit{Monitor}. \textit{Gateway} allocates a specific channel to pass these requests and isolate them from external requests.

A service provider can manage numerous services that are distributed in many networks. And the physical addresses of these services may not be static. Therefore, these requests only contain service identifiers assigned to running services. To map service identifiers to their latest IP addresses, \textit{Name Service} is required.

\subsubsection{Name Service}
\label{sec:name_service}
\textit{Name Service} is used to identify service consumers and services. It provides various query functionalities for querying identity information of service consumers such as DIDs, public keys, public service identifiers, and meta-information about their real identities. Typical usage is for the verification of the existence of service consumers. Besides, it serves the \textit{Gateway} to support queries of service information such as service IP addresses based on service identifiers.

\subsubsection{Monitor}
As a service platform, both service providers and consumers have demands of monitoring running services for the purposes such as data analysis and exception catching. \textit{Monitor} provides functionalities of collecting service data such as requests and status of nodes in specific services. These data can also be used for security audits, especially regarding the use of \textit{Permissioned Pattern}.

\section{Selfid}
Selfid is an SSIaaS platform with full-fledged features and functionalities illustrated in Section~\ref{sec:ssiaas}. As a platform, Selfid provides full support for service providers to manage, maintain, and extend all types of functionalities. Although the implementations of different modules vary in programming languages and tech stacks (e.g., \textit{Parser} implemented in Java, \textit{Gateway} in C++, \textit{Name Service} in Rust, and \textit{Monitor} in Go), all modules concretize our proposed service protocols with the same specification of data exchange and communication protocols. Furthermore, Selfid provides unified JSON-style application programming interfaces (APIs) for service consumers. Therefore, the whole platform can be regarded as an interoperable service that can be integrated into existing systems to provide flexible and scalable SSI-based functionalities.

For each SSI service, we define common interfaces and generalized protocols for all components to make them interoperable and independent of their internal architectures. Besides, all architectural patterns formulated in Section~\ref{sec:pattern} are unified in our SSIaaS platform to enable customizability. In this manner, service consumers can individualize their services and adopt optimal combinations without considering implementation details and integration compatibility. Selfid also allows service consumers to transfer data from an existing service to another with a different service architecture while preserving SSI-related functionalities. In principle, each service is assigned to a unique network location where its server applications are deployed to a Docker container after being built by \textit{Parser}. Consequently, services are relatively isolated from each other but uniformly managed by a container orchestration system (Kubernetes in our case). With the support of \textit{Gateway} and \textit{Name Service}, services are also portable and independent of their physical environments.

Particularly, we adopt Ethereum as the underlying distributed ledger in the \textit{Permissionless Pattern} and implement a collection of smart contracts in Solidity. In \textit{Permissioned Pattern}, both CDL and PDL are developed based on the HyperLedger Fabric with customizable consensus mechanisms.

\section{Discussion}
We discuss our SSIaaS architecture with respect to interoperability, customizability, and scalability.

\paragraph{Interoperability}
Identities registered in an SSI service are globally usable and can cross various system implementations. Besides, each SSI service consists of four relatively independent components that are low in coupling and high in cohesion. As illustrated in Section~\ref{sec:service_architecture}, components communicate with each other without the requirement of obtaining the internal status of the target. We demonstrate syntactic interoperability by implementing common interfaces via JSON-style APIs in Selfid. Furthermore, semantic interoperability is also achieved in our architecture, which can be seen by the generalized protocols in Section~\ref{sec:service_architecture} and service management modules in Section~\ref{sec:management}.

\paragraph{Customizability}
As shown in Section~\ref{sec:pattern}, a collection of architectural patterns are available for service consumers to customize their services. Thanks to interoperability, architectural decisions can adapt customized SSI services to specific scenarios without changing SSI functionalities.

\paragraph{Scalability}
Theoretically, an SSI service has better scalability than centralized identity services due to the nonnecessity of storing and maintaining identities controlled by users. As the number of users increases, the cost should at most increase linearly. But in practice, additional factors need to be taken into consideration. In our SSIaaS architecture, the \textit{Data Component} dominates the cost owning to the adoption of the DLT. Therefore, the scalability is also significantly affected by the consensus mechanism of distributed ledgers, especially in the \textit{Permissionless Pattern} and \textit{Sub-Ledger Pattern}.

\section{Conclusion}
\label{sec:conclusion}
As a promising identity model, SSI is changing the way of digital identity management and well suits Web3 \cite{voshmgir_token_2020}. By combining with the DLT, SSI systems present attractive advantages such as privacy protection, security enhancement, high availability.

In this paper, we have formulated the self-sovereign identity as a service to facilitate the interoperable, customizable, and scalable use of the DLT-based SSI in practice. We have also practicalized SSIaaS with a novel architecture by elaborating the service concept, SSI, and DLT. Besides, we have proposed an architecture for SSI services and formulated a collection of architectural patterns for SSI service components. For each component, we have provided evaluations for its architectural patterns as the insights of making architectural decisions. Selfied, an SSIaaS platform, has been developed based on our architecture to demonstrate the feasibility in practice. Furthermore, we have discussed the interoperability, customizability, and scalability of our architecture.

Our next research direction is to conduct an in-depth security analysis for SSIaaS and extract more architectural patterns for SSI service components. We also plan to explore new methods of practicalizing the DLT-based SSI.

\bibliographystyle{IEEEtran}
\bibliography{IEEEabrv,bibtex/bib/references}

\end{document}